\documentclass[runningheads]{llncs}
\usepackage{color}
\usepackage{graphicx}
\usepackage{algorithm}
\usepackage{algorithmicx}
\usepackage{algpseudocode}
\usepackage{amsmath}
\usepackage{amsfonts}
\usepackage{float}
\usepackage{booktabs}
\PassOptionsToPackage{hyphens}{url}
\usepackage{authblk}
\usepackage{bbding}
\usepackage[marginal]{footmisc}

\begin{document}
\title{SFE-GACN: A Novel Unknown Attack Detection under Insufficient Data via Intra Categories Generation in Embedding Space}
\titlerunning{SFE-GACN}
%
\author{Ao Liu,
Yunpeng Wang \Envelope,
Tao Li
}

\authorrunning{Ao Liu et al.}
%
\institute{College of cybersecurity, Sichuan University, Chengdu, 610065 China \\ \email{wyp@scu.edu.cn}
}

\maketitle
\begin{abstract}
\vspace{-0.6cm}
In the network traffic intrusion detection, deep learning based schemes have attracted lots of achievements. However, in real-world scenarios, data is often insufficient (few-shot), which leads to various deviations between  the models prediction and the ground truth. Consequently, downstream tasks such as unknown attack detection based on few-shot will be limited by insufficient data. In this paper, we propose a novel unknown attack detection method based on Intra Categories Generation in Embedding Space, namely SFE-GACN, which might be the solution of few-shot problem. Concretely, we first propose Session Feature Embedding (SFE) to summarize the context of basic granularity of network traffic: sessions, bring the insufficient data to the pre-trained embedding space. In this way, we achieve the goal of preliminary information extension in the few-shot case. Second, we further propose the Generative Adversarial Cooperative Network (GACN), which improves the conventional Generative Adversarial Network by supervising the generated sample to avoid falling into similar categories, and thus enables samples to generate intra categories. Our proposed SFE-GACN achieved that it can accurately generate session samples in the case of few-shot, and ensure the difference between categories during data augmentation. The detection results show that compared to the state-of-the-art method, the average TPR is 8.38\% higher, and the average FPR is 12.77\% lower. In addition, we evaluated the graphics generation capabilities of GACN on the graphics dataset, the result shows our proposed GACN can be popularized for generating easy-confused multi-categories graphics.

\keywords{Session Feature Embedding  \and Generative Adversarial Cooperative Network (GACN) \and Few-Shot \and Unknown Attack Detection \and Intra Categories Generation.}
\end{abstract}
\section{Introduction}
Network intrusion detection (ID) is a hotspot benefitting from the development of machine learning. Most of studies generalize feature sets from network traffic as the basis for further detection, represents session as tensor through training classifiers based on labeled datasets, and then finds behavioral characteristics of suspicious attacks. Benefitting from efficient machine learning tools, the detection task is transformed into a learning task on the feature sets by utilizing the classification model, such as deep networks with high computing power. 

\begin{figure}[htb]
\centering
\includegraphics[width=10.38cm, height=4.2cm]{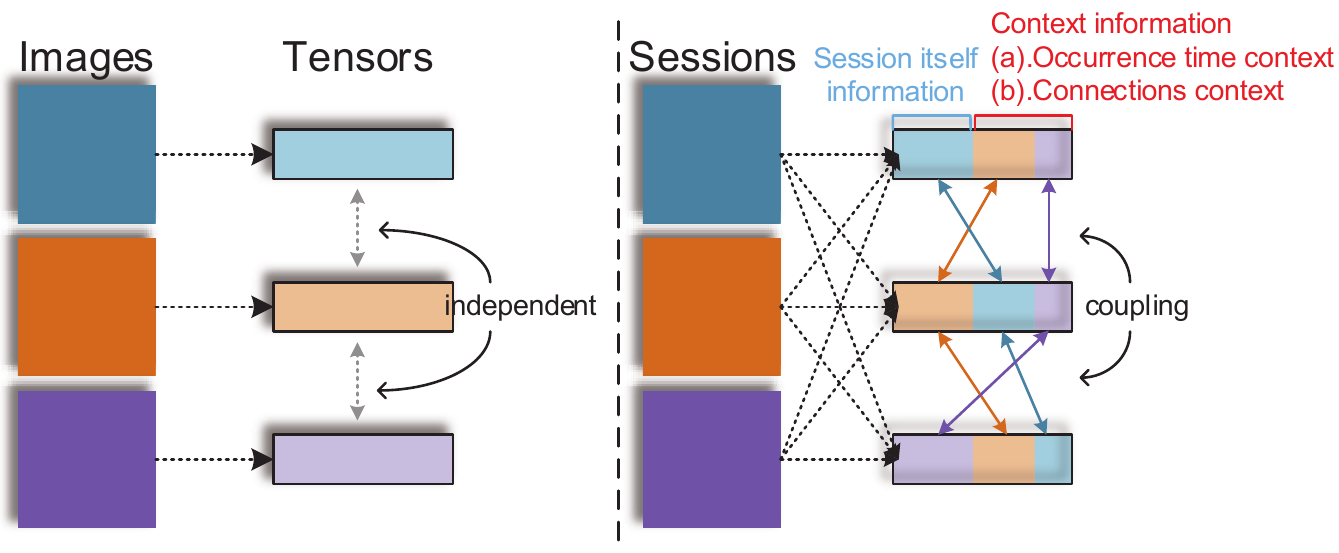}
\caption{Different coupling phenomena between graphics date set and session feature set} \label{fig1}
\end{figure}

However, adequate data cannot be guaranteed in most practical scenarios, conflicts emerge between data-hungry models and data-insufficient application scenarios. Further, downstream tasks such as unknown attack detection under few-shot prior information will be greatly influenced. These difficulties are encountered when seeking relevant research: ID is a topic restricted by application scenarios. In the related fields such as graphic classification, there are extensive researches on few-shot learning and unknown sample detection. However, session samples are coupled which is described in Fig.~\ref{fig1}. and this phenomenon does not exist in the graphic samples, so we cannot directly use the research results of graphic classification field. The specific difficulties of unknown attack detection under few-shot are as follows:

\begin{itemize}

\item {\bf Difficulty in finding the trade-off between model depth and data volume.} Intrusion traffic with sufficient data can be accurately detected, but this is not a common situation. For some subdivision tasks such as detection under insufficient data, the deep model cannot be fully trained, while the shallow model cannot fully fit feature sets. Concretely, the current state-of-the-art methods such as~\cite{1} requires a lot of prior knowledge, which is not satisfied in the scenario targeted by this article. So, we need to design a framework for detection tasks that is more subdivided.

\item {\bf Imperfection of data augmentation method.} Due to the fragmentation of application scenarios, model-based or metrics-based few-shot methods~\cite{2} cannot be used directly. Therefore, data augmentation for insufficient data is the solution to few-shot. Among them, Generative Adversarial Network (GAN)~\cite{3} is one of the most widely used methods~\cite{4}. However, GAN cannot guarantee the deviation between easily confused categories, that is, the GAN can only guarantee the similarity between the generated sample and the target category instead of guaranteeing the deviation between the generated sample and the similar categories sample.

\end{itemize}

It can be seen from these difficulties that unknown attack detection under few-shot is a comprehensive problem. So, we designed a full stack method to solve these problems comprehensively. We first propose SFE to summarize the context of session features, then propose GACN to implement intra categories generation, and finally improve the unknown attack detection method for the detection task. Our proposed method has the following advantages:

\begin{itemize}

\item {\bf Insufficient samples will be augmented by prior knowledge.} We propose a method for embedding session features to decouple the sessions to make them independent from the session context, bringing prior contextual information to the target sample. Further, through the pre-trained embedded model, few-shot traffic information will be augmented through prior knowledge, which can initially settle the few-shot problem. 

\item {\bf Generated samples will not be confused.} We propose GACN to solve the problem of confusing generated samples. Compared with only GAN~\cite{5}, the adversarial generated samples will be constrained by the cooperation model, to guarantee deviation between generated samples and similar categories samples.

\item {\bf More customized unknown attack detection method.}  Based on the proposals of SFE and GACN, we obtain the accurately augmented traffic data. Furthermore, we improve the existing unknown traffic detection method RTC~\cite{1}, to make it more suitable for unknown attack detection scenarios under few-shot.

\end{itemize}

Therefore, this paper comprehensively considers two factors and proposes a solution to solve them simultaneously. The method we proposed can mine unknown attacks that occur in the type of traffic to be detected in a scenario with fewer prior samples. 

{\color{blue}Specifically, in Section 3, we will introduce the SFE-GACN in detail: Section 3.1 for SFE; Section 3.2 for GACN; in Section 3.3 for improved RTC. In Section 4, we will evaluate our method in detail, in Section 4.1 we will evaluate the effectiveness of SFE, in Section 4.2 we will evaluate the effectiveness of GACN, in Section 4.3 we will carry out unknown attack detection experiments under insufficient data, and give the comparison results with the current state-of-the-art methods.}

\section{Related Work}
Few researches on intrusion detection have considered both few-shot and unknown attack detection concurrently, and therefore we will discuss the related literature separately.

{\color{blue}\textbf{Few-Shot Learning.}} Considering the scenarios for IDSs, when prior knowledge or target data is insuffi-cient, large-scale machine learning tools such as deep neural networks will not be adequately trained. Xian et al.~\cite{6} used three conditional GANs to generate embedding features step by step. {\color{blue}In their proposed method, the feature of target domain is taken as the object of claim, but the boundary problem of easily confused feature domain is not discussed}. Yong et al.~\cite{7} combined GAN and VAE, constrained the input of gen-erator to VAE, and improved generation accuracy. {\color{blue}But its task is to generate single category samples, so it cannot to migrate to multi categories intrusion detection.} Schonfeld et al. ~\cite{8} used two VAEs of the same structure, one to encode the image and the other to decode the class embedding. {\color{blue}The method mainly aims at the generalization problem of the model. While generating multi class samples, the generalization constraints added by the model will be gradually blurred with the introduction of Gaussian noise, which will let generated samples falling into similar categories.} Annadani et al. ~\cite{9} proved that introducing semantics into embed-ding space is beneficial to Few-Shot learning, {\color{blue}but it is not suitable for multi-categories and easily-confused intrusion traffic samples.} Kodirov et al.~\cite{10} used a semantic self-encoder to realize zero shot learning, which solved the problem of domain shift of training set and test set to a certain extent. {\color{blue}To some extent, this method provides us with the idea of human intervention in the semantic embedding space (supervised learning by fitting ground truth). However, because it still takes the initial sample as the fitting direction of convergence, it is unable to add "clear boundary" as the constraint condition, and further, it is still unable to generate accurate "within boundary samples" in the semantic embedding space, that is, intra-categories samples.} The work of ~\cite{11} proposed two methods, Deep-RIS and Deep-RULE, to solve the problem in different few-shot situations. IDSs can provide a certain level of protection to computer networks. {\color{blue}The unsupervised learning introduced by this method can obtain more accurate semantic knowledge in the embedded space, but on the other hand, the distribution of easily confused categories in the embedded space presents entanglement (verified in Section 4.1), so the unsupervised learning based on similarity will introduce more bias to the model}. 

{\color{blue}\textbf{Unknown attack detection.}} ~\cite{12} proposed a probabilistic approach and implements a prototype system ZePro for zero-day attack path identification. {\color{blue}However, this method is based on a large number of attack information, and the problem of intrusion detection of encrypted traffic only has a small amount of flow information.} Duessel et al. ~\cite{13} presented a new data representation diagram that allows us to integrate syntactic and sequential features of payloads in a unified feature space, provided a great solution for context-aware intrusions detection. {\color{blue}However, this method does not consider the migration of existing methods of low shot learning to achieve higher accuracy}. Zhang et al.~\cite{14,15} took the first step toward formally modeling network diversity as a security metric by designing and evaluating a series of diversity metrics. {\color{blue}However, although this method expands the diversity, it ignores the establishment of single category feature learning, that is to say, the method can not deal with easily confused intrusion samples}. The work of~\cite{16} designed heuristic algorithms to estimate the network attack surface while reducing the effort spent on calculating attack surface for individual resources. {\color{blue}However, this method also needs a lot of network attack information, which is not in accordance with the requirements of encrypted traffic intrusion detection task.} Zhang et al. ~\cite{1} proposed a new scheme of Robust statistical Trafﬁc Classiﬁcation (RTC) by combining supervised and unsupervised machine learning techniques to meet the challenge of unknown network trafﬁc classiﬁcation. However, if this meth-od is directly applied in attack detection, it will cause a large false positive rate. The reason is as follows. First, the shallow model used by RTC is not sufficient to fit the session feature set. Second, in RTC, the method of judging clusters during clustering is too simple to be extended. So, in this paper, we will improve the RTC to make it more suitable for unknown attack detection. We designed a category classification method for a single cluster, using a deep model instead of a shallow model, and finally reducing the false positive rate.

\section{SFE-GACN: The Framework of  Unknown Attack Detection}
\subsection{Session Features Embedding}
In natural language processing, words in a sentence are mapped to several vectors with independent features through word embedding~\cite{20}, which is no longer dependent on sentences. We take it as reference, and customize Session Feature Embedding method in order to reduce coupling between samples. The specific process is shown in Algorithm~\ref{alg:Session Features Embedding}.

\begin{algorithm}[htb]
        \caption{Session Features Embedding}
        \label{alg:Session Features Embedding}
        \begin{algorithmic}[1] 
            \Require Session features $\mathcal{F}$, embedding dimension $N$, window of embedding $c$
            \For{$i = 1$ to $\mathit{M}$}
            \State $v_i \gets 
            \mathrm{Binary}(\emph{max}
            \{f_{1i},\ldots,f_{ti}\})$ 
            \State $W^{(1)}_i \gets 
            \mathrm{Random\ Initialization}(W_{v_{i}N})$ 
            \State $W^{(2)}_i \gets 
            \mathrm{Random\ Initialization}(W_{N v_{i}})$ 
            \State $\mathcal{L} \gets \mathrm{Binary}(
            (f_{1i},\ldots,f_{ti})^T, bit_{max})$
            \State $X_k \gets \emptyset$ 
            \State $y_k \gets \emptyset$ 
            \For{$j = c$ to $t$}
                \State $X_k \gets X_k \bigcup \left(
                \sum _ { p = j - c } ^ { c - 1 } \mathcal { L } _ { p } + \sum _ { p = c + 1 } ^ { j + c } \mathcal { L } _ { p }\right) $
                \State $ y_k \gets \mathcal{L}_c $
            \EndFor
            \State $\mathrm{Model} \gets \mathrm{Sigmoid}\left(
            \mathrm{Linear}\left(x W_i^{(1)}\right)W_i^{(2)}\right)$
            \State $y^{output} \gets \mathrm{Model}(X_k)$
            \State $\Theta_i^{(l)} \gets  \Theta_i^{(l)} +
            \eta \nabla \left( \sum y _ { k } \log y ^ {output}  + \sum \left( 1 - y _ { k } \right) \log \left( 1 - y ^ { output } \right) \right)$
            \State $E_i \gets \mathcal { L }W_i^{(1)} $
        \EndFor
        \State $E \gets \left( E_1, \ldots, E_m \right)$
        \Ensure $E$
        \end{algorithmic}
    \end{algorithm}

Algorithm~\ref{alg:Session Features Embedding} presents the proposed method of Session Features Embedding, given a feature set $\mathcal{F}$ to obtain its embedded feature set $E$. First, binary transformation~\cite{21} is used to convert session features to binary representation. Since different features in the sample have inconsistent data type and data scale, set a maximum bit set $bit_{max} = \left\{v_1,\ldots,v_M\right\}$ based on the maximum number of the binary code of each feature. Convert different features to integer type, and finally map them to a sequence of 0 and 1, In order to keep the uniform coding length, fill the maximum bits with 0. After that, the session features are mapped int $0-1$ sequences of different lengths. We synthesize these small sequences into large $0-1$ sequences and average the weight of each feature. This process is completed by column embedding, which will be elaborated on next paragraph.

\begin{figure}
\centering
\includegraphics[width=7.09cm, height=8.4cm]{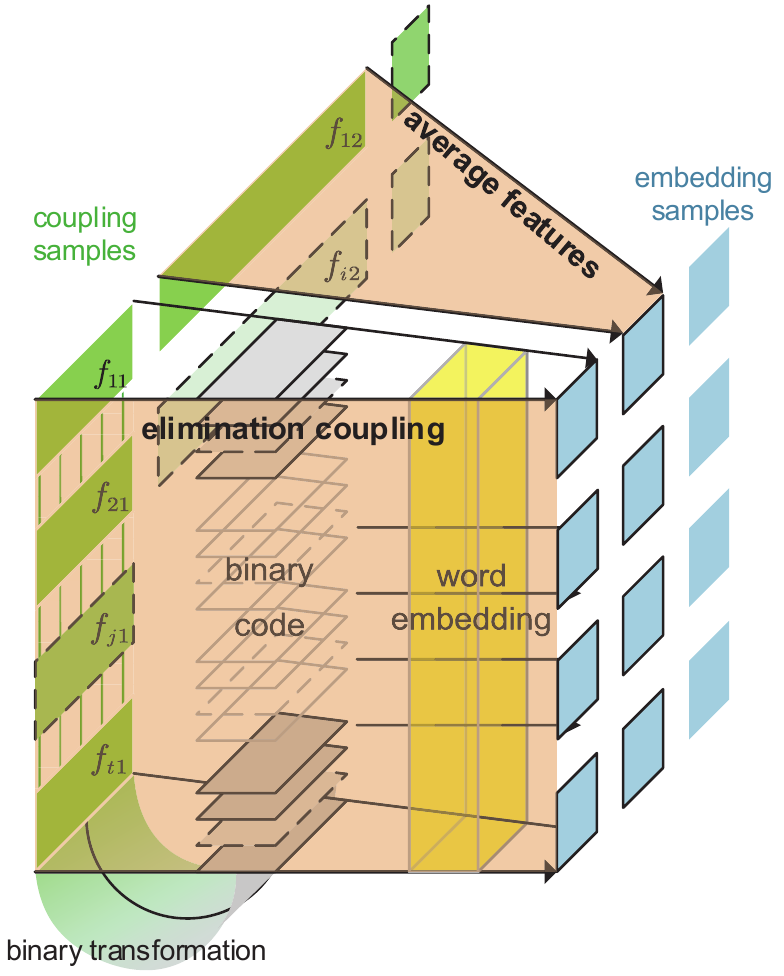}
\caption{Process of Session Features Embedding} \label{fig2}
\end{figure}

In order to get the embedding space of the sample set itself, we regard the column vector of each feature in sample set $\mathcal { F } = \left( \begin{array} { c c c } f _ { 11 } & \cdots & f _ { 1 M } \\ \vdots & \ddots & \vdots \\ f _ { t 1 } & \cdots & f _ { t M } \end{array} \right)$ as a sentence set.$\left\{f_{i1},\ldots,f_{iM}\right\}$ is all features of one sample, that is, there are altogether M sentences $S=\left\{s_1,\ldots,s_M\right\}$,$s_i=\left\{f_{1i},\ldots,f_{ti}\right\}$, and each word in $s_i$ is represented by binary representation. For $s_i$, we traverse each word and use its contextual information to predict it, so we can get the embedding vector of each word. Specifically, we build two trainable matrices $\Theta ^ { ( l ) } = \left( W_{v_i N}^{( 1 )},W_{Nv_i}^{(2)}\right)$ where the output dimension of the first layer is $N$, and the output dimension of the second layer is set to $v_i$. Then we use Stochastic Gradient Descent (SGD) to update the weight of $\Theta^{(l)}$ through back propagation. In order to obtain the embedding vector, we use the first matrix $W_{v_iN}^{(1)}$ to transform the binary encoded sentence of $t\times v_i$ into a vector of $t\times N(N<v_i)$. The algorithm is implemented for each $s_i\in S$, and then the resulting set of embedding vectors is vertically merged to obtain the total embedding matrix $E_{t\times \sum{bit_{max}}}$, each row contains the total vector of all the features of the samples after embedding. The process is shown in Fig.~\ref{fig2}.

\subsection{Generative Adversarial-Cooperative Network}
\begin{figure}[tb]
\centering
\includegraphics[width=5.36cm, height=6.18cm]{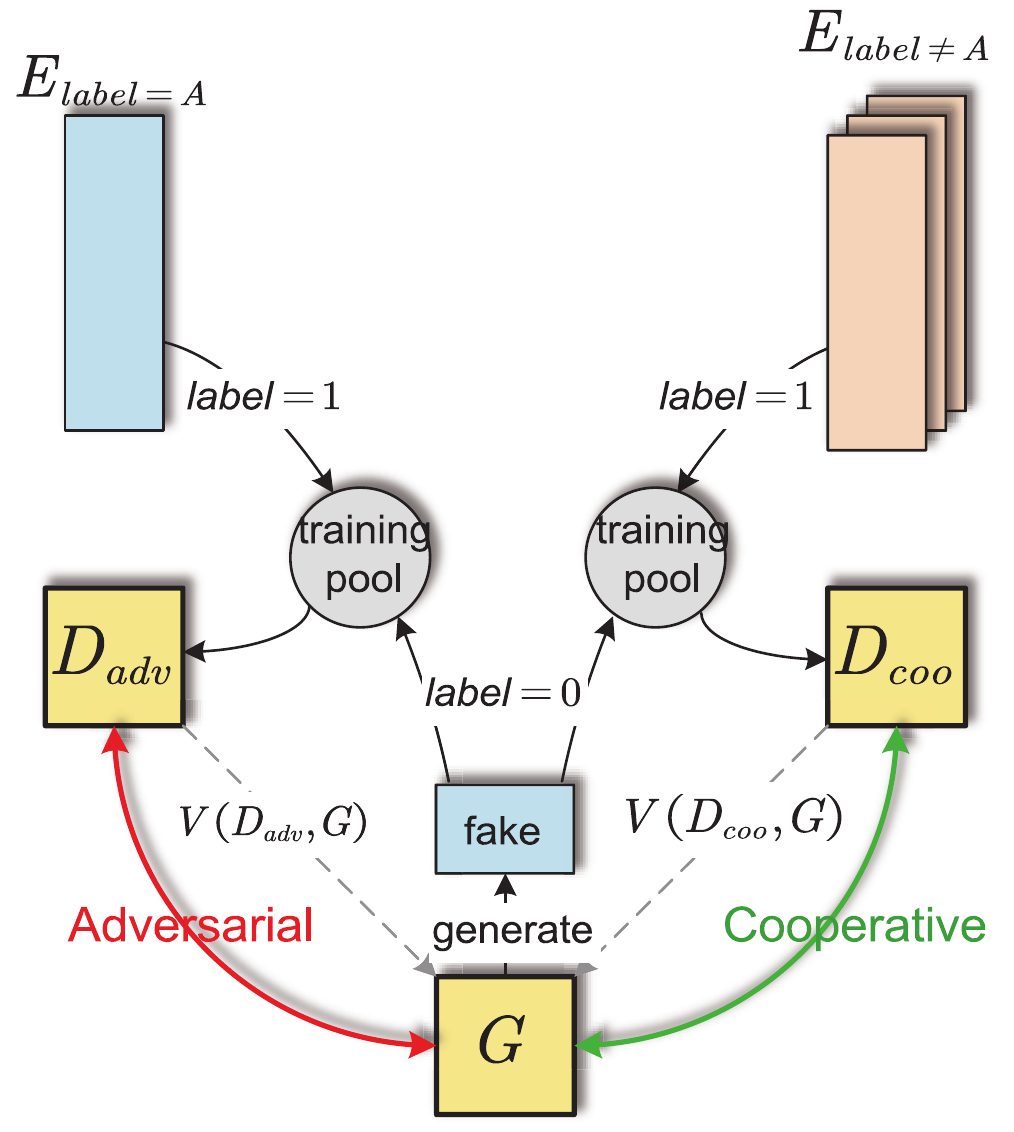}
\caption{The relationship among three models of GACN} \label{fig3}
\end{figure} 
\begin{figure}[tb]
\centering
\includegraphics[width=11.49cm, height=5.31cm]{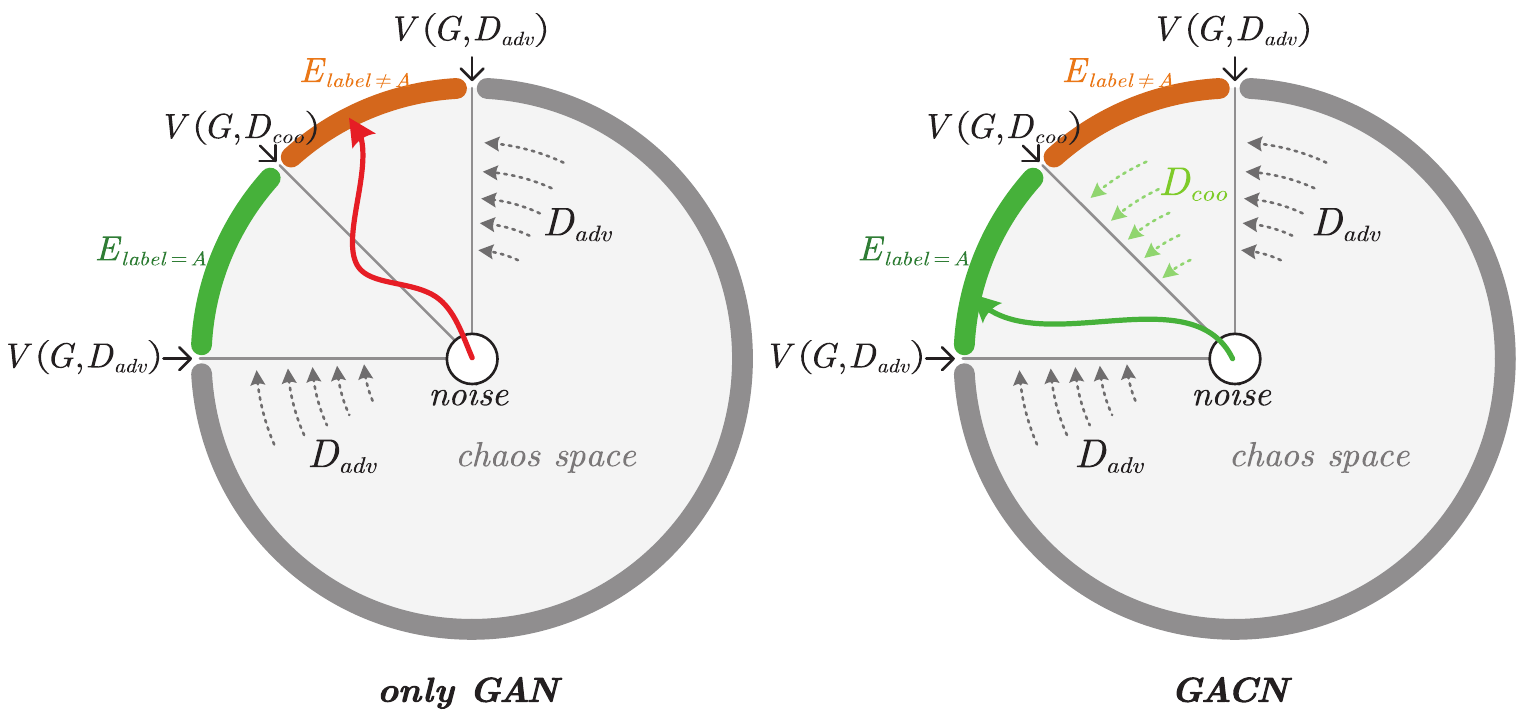}
\caption{The process of generating samples for only GAN and GACN respectively} \label{fig4}
\end{figure}
\noindent
We hope to conduct data augmentation for each category of samples in $E$ separately rather than uniformly with the GACN, the specific method is presented by Algorithm~\ref{algorithm2}. We firstly generate each kind of sample effectively while maintaining the deviation between the generated samples and the other label samples, which are called side samples. The core of GACN is to use $D_{coo}$ (Discriminator Cooperative) to supervise the training direction of $G$ (Generator) when $G$ and $D_{adv}$ (Discriminator Adversarial) are engaged in adversarial training. While $G$ gradually fits the generated sample space $E_{label=A}$, it can avoid moving forward to the sample space of other labels $E_{label\neq A}$  by adjusting its gradient descent degree. 
\par At the beginning of GACN, we manually specific some parameters and initialize three deep neural network models: $D_{adv}$ is to monitor whether the generated samples fit the target category $label=A$, $D_{coo}$ is used to monitor whether the generated sample is close to the side category $label\neq A$, and $G$ is used to generate the samples, respectively. First, we train $k$ rounds in $D_{adv}$ and $D_{coo}$ normally, then we train $G$. During this process, GACN can constantly distinguish if the samples generated by $G$ will incline to the side category. If they do, $G$ is rolled back until the samples generated by $G$ no longer have side category characteristics at all. Specifically, during process of using $D_{coo}$ to predict the generated samples, if the output value of the last sigmoid function approaches 0.5 and does not converge significantly anymore (which means that $G$ has fallen into the sample space of the side category), is able to roll back $G$ to the previous state, and negate the gradient descent direction that causes generating samples of $label\neq A$. Then the new weight of $G$ is calculated by SGD, through repeated negation and renewal, the direction of gradient descent will be affected by $D_{adv}$ and $D_{coo}$ at the same time, to ensure the samples generated by $G$ is no longer judged as $label\neq A$ by $D_{coo}$. In order to compensate for the multi-model training gap caused by rollback, $D_{coo}$ will be retrained after rollback so that the three models can work against/cooperate with each other to promote the positive iteration of $G$.
\par As shown in Fig.~\ref{fig3}, where are three networks in the GACN: $G$, $D_{adv}$, and $D_{coo}$. $G$ and $D_{adv}$ are adversarial, while $G$ and $D_{coo}$ are cooperative. In the process of adversarial training between $G$ and $D_{adv}$, real samples are gradually generated similarly, meanwhile under the supervision of $D_{coo}$, the generated samples are always kept distinct from $E_{label\neq A}$. The term of $V\left(D_{adv},G\right)$ represents the difference between the generated samples and the real samples. In the training process of GACN, the generated samples gradually approach the real sample space while maintaining a distinction from $E_{label\neq A}$. Furthermore, as shown in Fig.~\ref{fig4}, when we only use GAN, the samples generated by $G$ for $E_{label=A}$ will partially enter the sample space of $E_{label\neq A}$. GACN will try to avoid this situation; $D_{coo}$ will supervise the samples generated by $G$ and separate them from $E_{label\neq A}$.

\begin{algorithm}[ht]
        \caption{Generative Adversarial Cooperative Network}
        \label{algorithm2}
        \begin{algorithmic}[1] 
        \Require feature set $E$, target label $A$, weights of $D_{adv}$: $\theta_{adv}^{(D)}$, weights of $D_{coo}$: $\theta_{coo}^{(D)}$, weights of $G$: $\theta^{(G)}$, training epoch of $D_{adv}$: $k$, rollback check cycle: $cy_r$, backup cycle: $cy_b$, rollback coefficient: $c_r$
        \For{number of training iterations}
            \For{$epoch$ = $1$ to $k$}
                \State $\mathrm{Randomly\ generate\ noise\ } Z_1$
                \State $X_{real} \gets \mathrm{sample} \left(E_{label=A }\right)$
                \State $X_{side}^{(1)} \gets \mathrm{sample} \left(E_{label \neq A }\right)$
                \State $img_{fake}^{(1)} \gets G\left(Z_1\right)$
                \State Train $D_{adv}$ on batch, $X_{real}$'s label is set to $1$, $img_{fake}^{(1)}$'s label is set to 0 
                \State Train $D_{coo}$ on batch, $X_{side}$'s label is set to $1$, $img_{fake}^{(1)}$'s label is set to 0
        \EndFor
        \State $\mathrm{Randomly\ generate\ noise\ } Z_2$
        \State Record output of $D _ { \operatorname { coo } } \left( G \left( Z _ { 2 } \right) \right)$
        \If{$epoch$ reaches the end of $cy_b$ cycle}
            \State $\theta_{back}^{(G)} \gets \theta^{(G)}$ //backup $\theta^{(G)}$ as standby model
        \EndIf
        \If{$D _ { c o o } \left( G \left( z _ { 2 } ^ { ( 1 ) } \right) \right)$ does not drop in $e_r$ epochs}
            \State $\theta^{(G)} \gets \theta _ { back } ^ { ( G ) } + c _ { r } \left( \theta ^ { ( G ) } - \theta _ { back } ^ { ( G ) } \right)$ // Rollback $\theta ^ { ( G ) }$
            \State $\mathrm{Randomly\ generate\ noise\ } Z_3$
            \State $bat_{ab} \gets epoch \mod cy_b$ // Calculate the number of negated training epochs

            \State $X_{side}^{(2)} \gets \mathrm{sample} \left(E_{label \neq A }\right)$ //Simultaneous acquisition of fake and side  samples with $bat_{ab}$ capacity
            \State $img_{fake}^{(1)} \gets G\left(Z_3\right)$
            \State Train $D_{coo}$ on batch, $X_{side}^{(2)}$'s label is set to $1$, $img_{fake}^{(2)}$'s label is set to 0
        \EndIf
        \State $\mathrm{Randomly\ generate\ noise\ } Z_4$
        \State $\theta^{(G)} \gets \theta ^ { ( G ) } + \nabla _ { \theta ^ { ( G ) } } \frac { 1 } { r } \sum _ { i = 1 } ^ { r } \left[ \log \left( 1 - D _ { a d v } \left( G \left( Z _ { 4 } \right) \right) \right) \right]$ // update the $\theta^{(G)}$ by descending its stochastic gradient
    \EndFor
    \Ensure $\theta_{adv}^{(D)}$, $\theta^{(G)}$
    \end{algorithmic}
    \end{algorithm}
\subsection{Two-step Unknown Attack Detection}

\par When there are a small size of labeled samples and a large size of unknown samples that include unknown attack, we first use GACN to augment the known information by fusing the known samples and the unknown samples, carrying out preliminary detection by clustering, then using the deep neural network to reduce the false positive rate of final detection. Algorithm~\ref{algorithm3} presents the proposed method of Two-step Unknown Attack Mining.
\begin{figure}[htb]
\centering
\includegraphics[width=7.92cm, height=6.63cm]{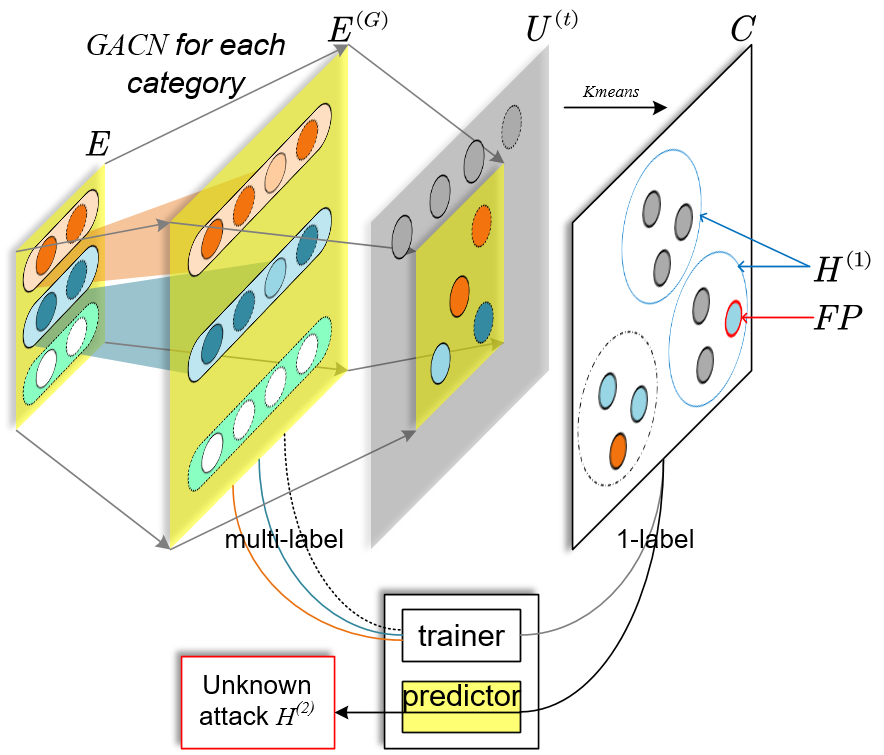}
\caption{The process of using $E$ to mine unknown attacks in $U$} \label{fig5}
\end{figure}
\par We use two steps method to mine the samples whose label never showed in $E$ from $U$. For the first step, augment $E$ to $E^{(G)}$ by GACN, integrate $E^{(G)}$ into unlabeled set $U$ to generate the total set $U^{(t)}$, and use KMeans to partition the $U^{(t)}$ into $q$ clusters $\left\{C_1,\ldots,C_q\right\}$. For each cluster, when the unlabeled samples are larger than a certain proportion $\delta$, all unlabeled samples are initially determined as unknown attack samples and expressed with $H^{(1)}$. There are a lot of false positive samples in $H^{(1)}$, so we solve it in the second step. Regard all samples in $H^{(1)}$ as unknown samples of a single class, after mixing a $q$-class set $E^{(G)}$  and 1-class set $H^{(1)}$ to generate a training set $K^{(X)}$, we train a $q+1$-class deep network $\mathbb{F}$ to classify $K^{(X)}$. We still use $H^{(1)}$ in the training set as the verification set, the purpose is to use $\mathbb{F}$ to eliminate the FP samples in $H^{(1)}$ and get a cleaner unknown attack sample set $H^{(2)}$. The process is shown in Fig.~\ref{fig5}.
\vspace{-0.4cm}
\begin{algorithm}[H]
        \caption{Two-step Unknown Attack Mining}
        \label{algorithm3}
        \begin{algorithmic}[1] 
        \Require $E$, Unknown label sessions feature set $U$, intra cluster decision rate $\delta$, known label list $Label$, length of Label: $K$
        \For{$i$ in $Label$}
            \State $E_{label=i}^{(G)} \gets \mathrm{GACN} \left(E_{label=i}\right)$
            \State $U^{(t)} \gets \mathrm{CONCAT}\left(U, E^{(G)}\right)$
        \EndFor
        \State $\underline{First step}$
        \State $H^{(1)} \gets \emptyset$ // First detection container
        \State obtain clusters $C = \left\{ C_1,\ldots,C_q\right\}$ with $\mathrm{KMeans}\left(U^{(t)}\right)$
        \For{$j = 1$ to $q$}
            \State $n_j^{(t)} \gets$ number of total samples in $C_j$
            \State $n_j^{(l)} \gets$ number of labeled samples in $C_j$
            \State $C_j^{(u)} \gets$ samples unlabeled in $C_j$
            \If{$n_j^{(l)} \leq \delta n_j^{(t)}$}
                \State $H^{(1)} \gets \left( H^{(1)} \bigcup C_j^{(u)} \right)$
            \EndIf
        \EndFor
        \State $\underline{Second step}$
        \State $H^{(2)} \gets \emptyset$ // Second detection container
        \State $K^{(X)} \gets \emptyset$ // Training set
        \State $K^{(y)} \gets \emptyset$ // Label of $K^{(X)}$
        \For{$i$ in $Label$}
            \State $K^{(X)} \gets K^{(X)} \bigcup E_{label=i}^{(G)}$
            \State $K^{(y)} \gets K^{(y)} \bigcup i$
        \EndFor
        \State $K^{(X)} \gets K^{(X)} \bigcup H^{(1)}$
        \State $K^{(y)} \gets K^{(y)} \bigcup unknown$ // Set the label of all samples in $H^{(1)}$ to $unknown$
        \State Training a neural network $\mathbb{F}$ as multi-class classifier by $\left(K^{(X)},K^{(y)}\right)$ for $\left(K+1\right)$ categories classification
        \For{$h_i^{(1)}$ in $H^{(1)}$}
            \If{$h_i^{(1)}$ is predicted as $unknown$}
                \State $H^{(2)} \gets H^{(2)} \bigcup h_i^{(1)}$
            \EndIf
        \EndFor
        \Ensure Sample set detected as unknown attack: $H^{(2)}$
        \end{algorithmic}
    \end{algorithm}

\section{Evaluation}
By using the matrix of pre training to get the embedded features under few-shot, SFE can represent the uncoupled features of samples to improve the detection accuracy. Based on SFE, GACN can get the augmented samples within the category. After that, the improved two-step method cooperates with the former two to complete the unknown attack detection under few-shot. This section verifies the effectiveness of SFE and GACN, and combines the multi-layer method to evaluate the detection indicators.
\subsection{Effectiveness of SFE}

We get CICIDS-2017~\cite{18} as the evaluation data set, Friday’s network traffic data in the dataset is obtained to train the embedding model $W_1^{(1)}$ then evaluate the effectiveness of SFE with other date traffic data. In this section. We only evaluate the classification performance of SFE to few-shot traffic, and the evaluation of unknown attack detection will be conducted in Section 4.3. 
\begin{algorithm}[htb]
    \caption{Point Walk}
    \label{algorithm4}
    \begin{algorithmic}[1] 
    \Require all embedding features $E^{(W)}$ and the size $N^{(E)}$ of them, statistical steps $w_s$, label set $\left\{l_1^{(E)},\ldots,l_{ce}^{(E)}\right\}$
    \State $E_{start}^{(w)} \gets$ randomly select a point from $E^{(w)}$
    \For{$i = 1$ to $ce$}
        \State $J_i \gets \emptyset$ // cycle end statistics container
        \State $J_i^{(in)} \gets \emptyset$ // intra cycle statistics container
    \EndFor
    \For{$j = 1$ to $N^{(E)}$}
        \If{$j=1$}
            $E_{step}^{(w)} \gets E_{start}^{(w)}$
        \Else
            \State $E_{step}^{(w)} \gets$ the nearest neighbor point of $E_{step}^{(w)}$
            \State $l_{step} \gets$ label of $E_{step}^{(w)}$
            \State $J_i^{(in)} \gets J_i^{(in)} \bigcup l_{step}$
        \EndIf
        \If{$j\mod w_s = 0$}
            \State count all labels in $J_i^{(in)}$, put them in corresponding $\left\{J_i,\ldots,J_{ce}\right\}$
            \State $J_i^{(in)} \gets \emptyset$
        \EndIf
    \EndFor
    \Ensure $\left\{J_i,\ldots,J_{ce}\right\}$
    
    \end{algorithmic}
    \end{algorithm}
\par
\begin{figure}[ht]
\centering
\includegraphics[width=8.77cm, height=6.53cm]{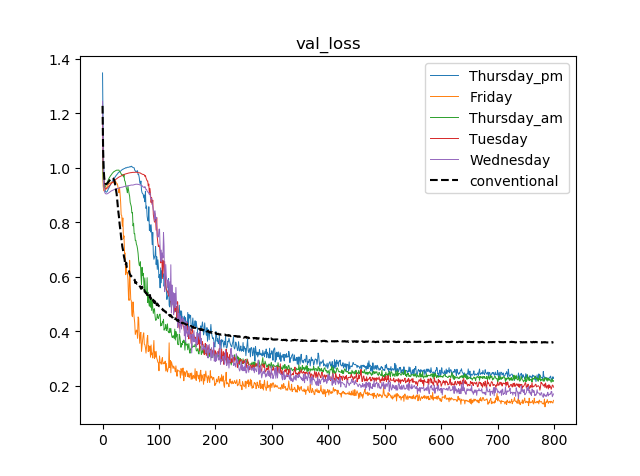}
\caption{Convergence of loss with embedded and conventional features in the case of few-shot samples, Friday's feature set is used to train $W_1^{(1)}$} \label{fig6}
\end{figure}

 In order to use the prior embedding matrix $W_1^{(1)}$  to process the unknown few-shot samples, and then get their positions in the embedded space, we reduce the sample size of traffic on other dates, and use $W_1^{(1)}$ to embed the traffic of few-shot samples to obtain the corresponding embedding features. Finally, a single-layer perceptron is used to train multiple classifiers, and the validation loss convergence of the classifiers is counted to evaluate the effectiveness of SFE.The normal traffic of all dates is reduced by 20 times, the attack traffic is reduced by 10 times, and the convergence is counted. The experimental results are shown in Fig.~\ref{fig6}. The lines of different colors in Fig.~\ref{fig6} represent the traffic data of different days, each of which contains part of the attack traffic.

As shown in Fig.~\ref{fig6}, the convergence rate of the conventional feature set is faster, but the final convergence rate is higher while the embedded feature converges to a lower value. Therefore, in the case of few-shot samples, using the pre-training embedded matrix can more accurately describe the sample characteristics.
\par We will continue to discuss how samples are distributed in the embedded space to evaluate the effect of SFE on sample coupling. The sample distribution in the embedded space is obtained by Point Walk, which is represented by Algorithm~\ref{algorithm4}. Point Walk starts from a random point and all samples are connected in series according to the nearest point. In this process, we count the number of different categories step-by-step in the window, and get the set of statistics $\left\{J_i,\ldots,J_{ce}\right\}$. For visualization, we take steps as the $X$-axis, and different samples in $J$ as the $Y$-axis. 

\begin{figure}[htb]
\centering
\includegraphics[width=11.55cm]{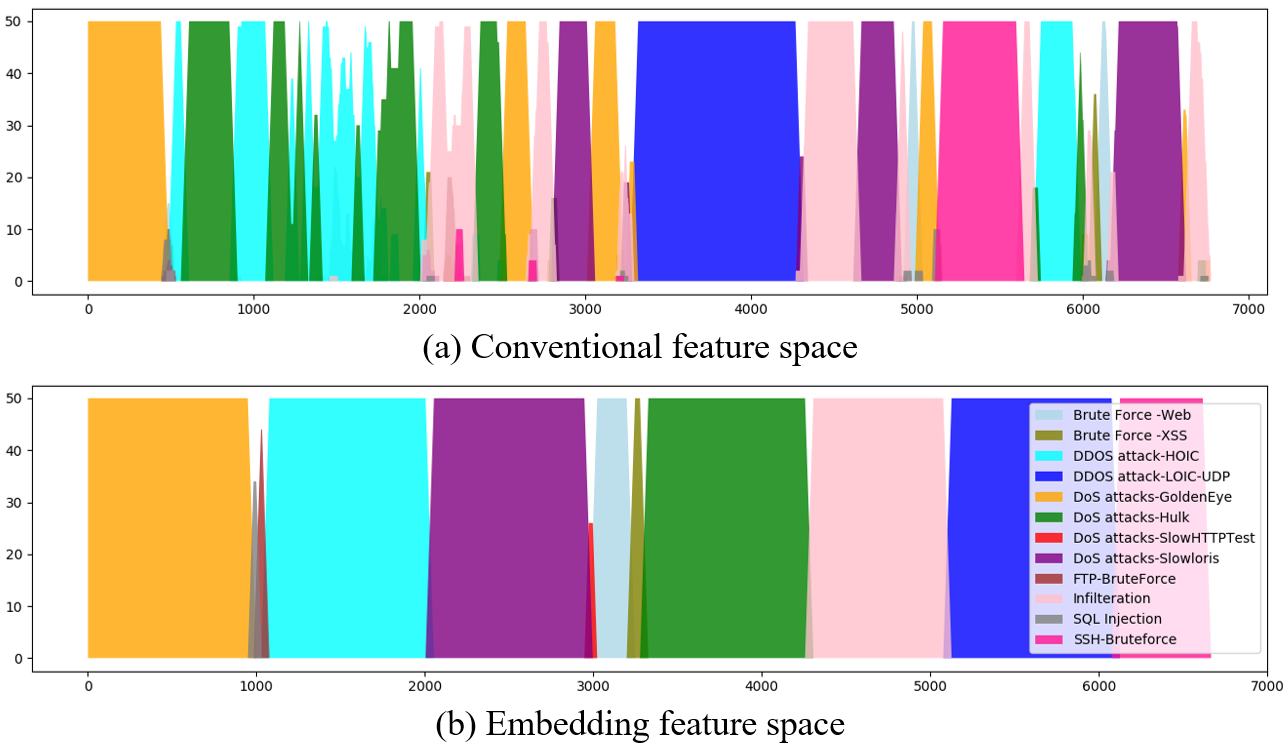}
\caption{The result of Point Walk in conventional feature space and  embedding space} \label{fig7}
\end{figure}

Using all the data from the IDS2018~\cite{18} dataset with richer categories, we retrain the embedded model and obtain the corresponding embedded dataset. Point Walk is used for the new embedded dataset to get the corresponding  $\left\{J_i,\ldots,J_{ce}\right\}$ and the coordinate map is drawn, the results are as shown in Fig.~\ref{fig7}.

In Fig.~\ref{fig7}, the Y-axis represents the category statistics within the sliding window as the point walks in the embedded space, and the different colors represent the attack samples of different categories. Fig.~\ref{fig7} shows that the sample categories experienced during Point Walk are always regular, and the Euclidean distances between samples in the same category are close. This is similar to the rule of word embedding~\cite{20}: the distance between ``apple'' and ``pear'' is much smaller than that between ``apple'' and ``Barack Obama'', which indicates that SFE successfully trains the embedded space of the samples so that the samples no longer rely solely on their traffic environment, and reduces the coupling between the samples.

\subsection{Effectiveness of GACN}
\begin{figure}[htb]
\centering
\includegraphics[width=7.84cm]{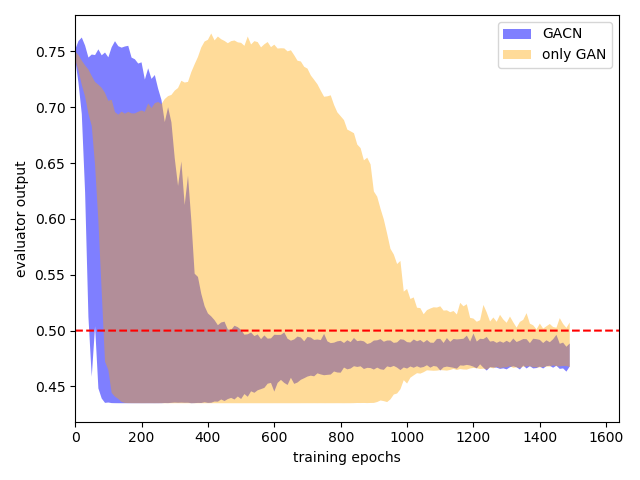}
\caption{Evaluate results of GACN and only GANl feature space} \label{fig8}
\end{figure}

In order to evaluate the effectiveness of GACN in preventing $G$ from inclining to side categories, we obtained a fashion-MNIST~\cite{19} data set that was more easily confused with different categories samples. We augment $E_{label=A}$ with $E_{label\neq A}$ as $X_{side}$, and at the same time, train an evaluator to determine if the generated samples in the iteration process are inclined to $X_{side}$. The results are shown in Fig.~\ref{fig8}, and the parameter settings of GACN and the models are shown in Table~\ref{tab1}. In order to more accurately evaluate the supervising ability of GACN, the initial random noise is set to $X_{side}+noise$.

During the pre-training process of the evaluator, we set the label of $E_{label=A}$  to 0 and the label of $X_{side}$ to 1. Therefore, when the output of the last layer of the evaluator (score) is less than 0.5, it is determined that the sample to be evaluated does not incline to $X_{side}$, otherwise, it means that it will happen with some results we did not expect: $G$ is moving towards the sample space of $X_{side}$. 

In Fig.~\ref{fig8}, the X-axis represents the training epoch of the GACN and only GAN; the upper bound represents the highest score given by the evaluator and the lower bound represents the lowest score. We find that with the increase of the number of iterations, the score of GACN quickly converges to less than 0.5, while the score of only GAN cannot converge for a long time, which means that the samples generated will be classified into $X_{side}$ by the evaluator, resulting in the deviation of the generated samples.

\begin{figure}[htb]
\centering
\includegraphics[width=11.55cm]{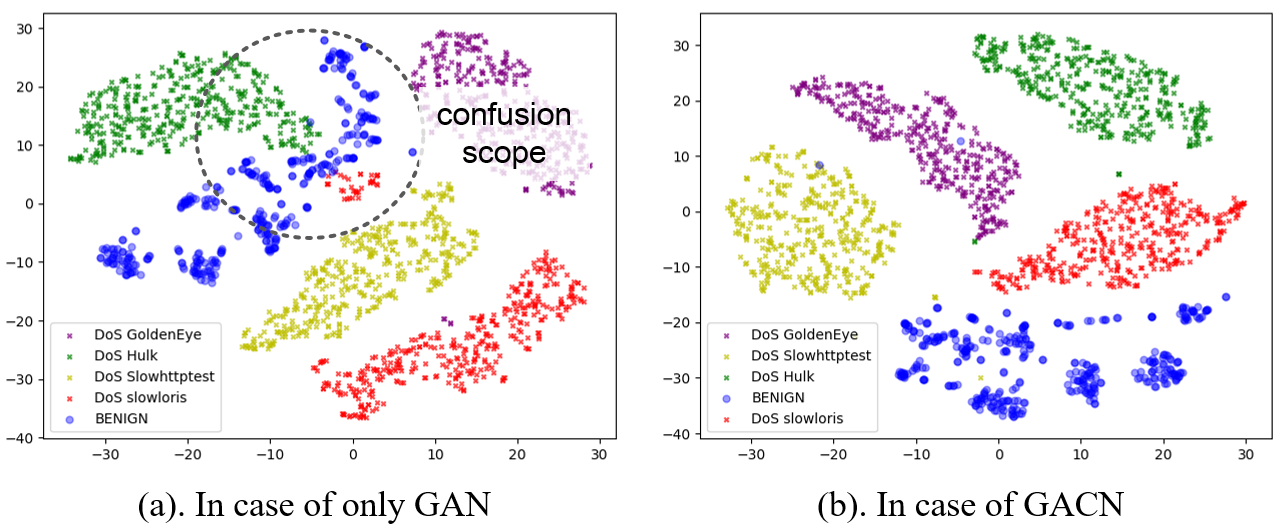}
\caption{Distribution of generated samples in embedding feature space} \label{fig9}
\end{figure}

In order to verify the distinction between different categories of samples generated by GACN at the session feature analyses topic, we use GACN for CICIDS-2017 data sets, then we apply t-SNE~\cite{22} to reduce the dimensions of features and visualize them. We resample $E$ to get $E^{(s)}$, so that each category without BENIGN is generated intra the class using GACN. For clarity of picture representation, we visualize the experimental results of using generated and BENIGN samples as shown in Fig.~\ref{fig9}. 

As can be seen from Fig.~\ref{fig9}, the boundary between samples generated by only GAN is fuzzy in confusion scope. With the increase of sample size, the classifier may misjudge. However, GACN always avoids generated samples close to the samples of other categories, so the boundary between the generated samples of different categories is obvious, which will improve the performance of the classifier.
\par Further, In order to evaluate the difference between GACN and only GAN in the traffic detection indicated, we use CICIDS-2017 embedded feature set $E$ as experimental data. We first randomly sampled $E$ for few-shot samples, and then kept the same sampling rate for each category of sample. We use GACN, only-GAN, and non-generate to change the sample size to augment data and obtaine $E^{(GACN)}$, $E^{(GAN)}$, $E^{(non)}$ , respectively. We then trained the deep network as multi-classifiers, and recorded the f1 score and time spending separately. The experimental results are shown in Table~\ref{tab1}. To reduce the error introduced by sampling, the experiment was repeated three times, and the average value of the result indicators was taken. The experiment used Nvdia K80 GPU as the training accelerator.
\begin{table}[htb]
\centering
\caption{Detection index in embedded space.}\label{tab1}
\begin{tabular}{llll}
\toprule
Indicators & $E^{(GACN)}$ & $E^{(GAN)}$ & $E^{(non)}$\\
\midrule
  f1 score & 0.9909 & 0.9714 & 0.9106 \\
  time overhead & 49.63s & 40.26s & 9.82s \\
\bottomrule
\end{tabular}
\end{table}
\par The experimental results show that in the case of few-shot samples, GACN increased the f1 score while maintaining a small increase in time overhead.
\subsection{Evaluation with Unknown Attack Detection}
We propose a multi-layer solution to solve the problem of unknown attack detection in few-shot samples. We extract 1/10 of each category attack by random sample from IDS2018 as a prior labeled feature set, and use it to mine unknown attacks in the remaining samples. The labeled and unlabeled scales of each category are shown in Table~\ref{tab2}. In order to reduce the calculation cost, attacks with large data size are reduced to 1/10 of their original size.
\par We test and evaluate each class of attacks in the unlabeled dataset as unknown attacks separately. When a category $c_u$ is regarded as an unknown attack, we delete the data in the labeled data set, then detect the $c_u$ samples in the pending detection samples. At the same time, RTC is used as the benchmark. The results are shown in Table~\ref{tab2}.
\par The experiments result show that our method improves TPR at some extent(increased by  8.38\%), as well as significantly reduced FPR(decreased by 12.77\%), which shows that our proposed method performs well.
\begin{table}[htb]
\centering
\caption{The scale of various categories of attacks in IDS2018 dataset and their detection TPR/FPR}
\begin{tabular}{lllll} 
\toprule
   Attack category & Labeled & Unlabeled & TPR(Ours/RTC) & FPR(Ours/RTC) \\
  \midrule
  Hulk	& 4619	& 41572	& 0.9576/0.8760	& 0.0293/0.1576 \\
  HTTPTest &	1398 &	12590 &	0.9302/0.8537 &	0.0249/0.1302 \\
  GoldenEye &	4151 &	37357 &	0.9403/0.8617 &	0.0211/0.1503 \\
  Slowloris &	1099 &	9891 &	0.9110/0.8276 &	0.0225/0.1698 \\
  FTP-BruteF &	1933 &	17402 &	0.9211/0.8164 &	0.0204/0.1211 \\
  SSH-BruteF &	1875 &	16883 &	0.9279/0.8593 &	0.0350/0.1979 \\
  HOIC &	6860 &	61741 &	0.9308/0.8390 &	0.0110/0.1308 \\
  \textbf{average} &	N/A &	N/A &	\textbf{0.9313/0.8475} &	\textbf{0.0234/0.1511} \\
  \bottomrule
  \end{tabular}
  \label{tab2}
\end{table}

\section{Conclusions and Future Work}
In this paper, we propose SFE-GACN as an unknown attack detection method under few-shot, which fills the gap in research in the target we aimed to investigate. It is based on the existing session feature set classification method. There are several advantages of SFE-GACN: 
\par (1) It can decouple the sessions in the feature set by embedding, and bring the prior information into the few-shot samples to complete the preliminary augmentation of few-shot samples. 
\par (2) When data augmentation is performed, samples in multiple categories are generated as intra categories to prevent confusion between generated samples. 
\par (3) It improves upon the conventional unknown attack detection methods, making it more suitable for detection under few-shot, and can be docked with SFE-GACN to complete the final detection task. 
\par SFE-GACN is used for the final detection task, which performance outperforms the current state-of-the-art method. However, there are also some points that need to be improved and extended in the future.
\par \textbf{(1) Optimal hyperparameter setting method in the model.} In practical applications, a large number of hyperparameters need to be customized by workers, including the time window and embedded dimensions in SFE, and the rollback judgment epoch, rollback coefficient, and backup cycle in GACN. The optimal selection method of these hyperparameters will be given in the future.
\par \textbf{(2) Universal scalability.} Even if we refine ID tasks to more targeted scenarios such as unknown detection tasks under few-shot, there are still some more detailed application scenarios to deal with. For example, multi classification or binary classification, how inadequate is the data, etc. We will continue to explore these specific scenarios in the future and expand the universal scalability of SFE-GACN.

\section*{Acknowledgments}
This work is partially sponsored by the National Science Foundation of China (U19A2068, U1736212).

\end{document}